\documentclass[aip,cha, reprint]{revtex4-2}

\usepackage{amssymb}
\usepackage{amsmath}
\usepackage{graphicx}% Include figure files
\usepackage{dcolumn}% Align table columns on decimal point
\usepackage{bm}% bold math
\usepackage[utf8]{inputenc}
\usepackage[T1]{fontenc}
\usepackage{mathptmx}
\usepackage{etoolbox}
\usepackage{xcolor,colortbl}
\usepackage{algorithm}
\usepackage{algpseudocode}
\usepackage{hyperref}

\makeatletter
\def\@email#1#2{%
 \endgroup
 \patchcmd{\titleblock@produce}
  {\frontmatter@RRAPformat}
  {\frontmatter@RRAPformat{\produce@RRAP{*#1\href{mailto:#2}{#2}}}\frontmatter@RRAPformat}
  {}{}
}%
\makeatother
\begin{document}

\preprint{AIP/123-QED}

\title{Unified Geometry-Guided ML-FTLE for Tracking Transient Chaos from Scalar Time Series}

\author{S. V. Manivelan}%
\affiliation{Department of Physics, M. R. Government Arts College (Affiliated to Bharathidasan University, Tiruchirappalli), Mannargudi, 614 001, Tamilnadu, India.}
\author{Andrei Velichko}%
\email[Corresponding Author: ]{velichkogf@gmail.com}
\affiliation{Institute of Physics and Technology, Petrozavodsk State University, Petrozavodsk, 185910, Russia.}
\author{I. Manimehan}%
\affiliation{Department of Physics, M. R. Government Arts College (Affiliated to Bharathidasan University, Tiruchirappalli), Mannargudi, 614 001, Tamilnadu, India.}
\date{\today}% It is always \today, today,
%  but any date may be explicitly specified

\begin{abstract}
Detecting transient chaos from scalar observations without governing equations represents a fundamental challenge in nonlinear dynamics. We propose a geometry-guided machine learning framework that unifies predictive trajectory divergence with macroscopic attractor morphology to track abrupt regime shifts. The methodology extracts a local instability scale via out-of-sample $k$-nearest neighbor forecast errors to establish the ML-FTLE estimator, subsequently mapping this temporal divergence onto a structural closeness matrix derived from a minimal dictionary of Poincar\'{e} occupancy grids. By employing partial least squares regression, we extract a latent geometric component calibrated directly to the empirical finite-time Lyapunov spectrum, yielding the Poincar\'{e}-based geometric-guided FTLE. Validation against analytical QR-FTLE baselines confirms that fusing topological state spaces with predictive divergence systematically improves continuous transition tracking. The Structural Similarity Index optimally resolves gradual damping, while Hausdorff Distance exhibits extreme resilience during abrupt phase-space collapses. Furthermore, macroscopic spatial discretization acts as a robust topological regularizer against additive Gaussian noise, preserving deterministic signatures even at moderate signal thresholds. This equation-free framework provides a highly accurate, noise-resilient diagnostic for monitoring structural transitions in complex non-stationary systems.
\end{abstract}

\maketitle

\begin{quotation}
Complex systems in nature, from shifting climate patterns to fluctuating financial markets, often undergo sudden regime shifts in behavior. A central challenge lies in detecting these transitions from empirical observations when only a single scalar measurement is accessible and the governing equations remain unknown. In this work, we demonstrate that transient chaos can be tracked from a single scalar signal by unifying two coordinated representations. The first leverages predictive divergence quantified by forecast error growth. The second utilizes attractor geometry captured by the deformation of delay coordinate occupancy maps. Fusing these perspectives produces an interpretable and robust Lyapunov scale indicator. The predictive machine learning estimator supplies the localized instability scale, while basis attractor geometry identifies the precise structural reorganization driving the regime collapse. Furthermore, rigorous validation against additive noise demonstrates that this macroscopic geometric representation preserves deterministic signatures and maintains high diagnostic accuracy even under substantial observational noise.
\end{quotation}
%\linenumbers

% \linenumbers\relax % removed for arXiv
\section{Introduction\label{Introduction}}
Quantifying instability from observational data is a central task in nonlinear dynamics, as many real systems exhibit transient chaotic regimes prior to settling into asymptotic periodic, quiescent, or fixed-point behaviour~\cite{lai2011transient}. In experimental and field applications, the governing equations of a system are often unavailable, and the only accessible data may be a single univariate time series. Under these constraints, finite-time Lyapunov exponents (FTLEs) serve as highly effective diagnostic quantities. Unlike global asymptotic Lyapunov exponents, FTLEs quantify the localized exponential stretching or contraction of trajectories over a finite observation window, providing a rigorous mathematical framework for tracking predictability loss, transient instability, and abrupt regime changes~\cite{Babaee2017ReducedOrder,Parlitz2016Estimating}.

When the governing equations of dynamical systems and their Jacobian are known, Lyapunov exponents and FTLEs can be computed from the tangent dynamics using standard orthonormalization procedures \cite{oseledec1968multiplicative}. However, this procedure becomes significantly more challenging in observational settings, where neither the full state vector nor the flow map is directly accessible. To address this, the approach relies on reconstructing a topologically equivalent phase space from a single scalar observable using delay coordinate embedding, grounded in Takens' theorem~\cite{Takens1981Detecting} and standard techniques in nonlinear time series analysis~\cite{Bradley2015Nonlinear}. Building upon this foundation, empirical algorithms estimate the maximal Lyapunov exponent by evaluating the local divergence of adjacent state vectors within the reconstructed phase space. The methodologies introduced by Wolf \emph{et al.}~\cite{Wolf1985Determining}, Rosenstein \emph{et al.}~\cite{Rosenstein1993A}, and Kantz~\cite{Kantz1994A} remain the standard references for this class of approaches.

While these classical estimators are foundational, their formulation is optimized for stationary or ergodic conditions and does not naturally yield a time resolved, window-by-window instability trajectory. Transient chaos presents a distinctly different challenge as the objective is not merely to determine whether the system is chaotic on average but to continuously track the temporal evolution of local instability and how the underlying attractor reorganizes during the transition. Furthermore, finite data length, measurement noise \cite{Yang2011A}, imperfect embedding \cite{Bradley2015Nonlinear}, and heterogeneous sampling density can all limit the reliability of Lyapunov estimates from scalar data, as emphasized in earlier work on the fundamental and practical limits of time series based exponent estimation \cite{Eckmann1992Fundamental}. Thus, a useful detector of transient chaos should combine a quantitative instability metric with an interpretable mapping of the topological deformations within phase space.

Recent machine learning approaches provide a route to data-driven stability analysis, including reservoir computing~\cite{Pathak2017Using}, neural predictors~\cite{Margazoglou2022Stability,Margazoglou2023DataDriven}, and deep learning models~\cite{MayoraCebollero2024Full} have been used to reproduce chaotic attractors, derive Lyapunov spectra, and approximate stability metrics directly from observed time series. Nevertheless, a purely predictive Lyapunov proxy does not fully solve the transient regime problem. Forecast error growth provides a useful numerical instability scale, but during an abrupt attractor collapse, it may exhibit temporal lag and fails to explain which geometric change caused the loss of chaotic behavior.

To obtain this missing structural information, a natural approach is to analyze the geometry of the reconstructed trajectory. Recurrence plots~\cite{Eckmann1987RecurrencePlots} and their quantification statistics~\cite{Marwan2007RecurrencePlots} provide a well-established toolkit for this purpose, linking topological features of the delay space representation~\cite{Iwanski1998RecurrencePlots} to regime type, parameter drift~\cite{Casdagli1997RecurrencePlotsRevisited}, stationarity loss~\cite{Manuca1996Stationarity}, and abrupt transitions. These results support a key premise that delay space geometry is not merely a visualization layer but a direct physical representation of dynamical structure. However, existing geometry-based diagnostics quantify either pairwise recurrences within a single window or statistics derived from individual reconstructed sets and they do not naturally yield a compact, time-resolved coordinate system that maps each window relative to a basis of representative attractor geometries~\cite{Kantz1994QuantifyingCloseness,Bradley2002RecurrencePlotsUPO}. This gap motivates the base attractor representation introduced here. Rather than comparing only neighboring windows, we construct a closeness matrix whose rows are time windows and whose columns are similarity scores to a small set of representative geometric regimes, turning evolving attractor morphology into a multivariate trajectory. Crucially, while highly sensitive to structural loss, geometric representations in isolation do not provide a calibrated Lyapunov scale quantity.

We therefore propose a geometry-guided ML-FTLE framework that fuses these two coordinated representations, namely predictive divergence and attractor geometry. The first component estimates a local Lyapunov scale signal, denoted $\hat{\lambda}_{\mathrm{ML}}$, from the slope of the log-transformed geometric mean absolute error (GMAE) of out-of-sample $k$-nearest neighbor forecasts across sliding windows. This explicitly extends the forecast divergence principle of Velichko \emph{et al.}~\cite{Velichko2025Novel}, establishing forecast error growth as a robust proxy for trajectory divergence in one-dimensional chaotic time series to a finite-time, window-resolved setting. The second component converts each window into a Poincar\'{e} like occupancy grid and compares it to a compact set of basis attractors, forming a low-dimensional geometric coordinate system for the evolving regime.

To map this structural representation onto a quantitative instability scale, Partial Least Squares Regression (PLSR)~\cite{Abdi2010Partial} is used as a supervised latent variable bridge~\cite{Zhu2020Latent}, extracting the single latent geometric component that maximally co-varies with $\hat{\lambda}_{\mathrm{ML}}$. Rather than treating forecast divergence and attractor morphology as separate detectors, we demonstrate that they encode complementary aspects of the same transition. ML-FTLE supplies the quantitative instability scale, basis attractor geometry identifies the structural phase space reorganization responsible for regime collapse, and their PLSR fusion yields an interpretable, noise-resilient Lyapunov scale indicator from a single scalar observable.

%The article is organized as follows. Section~\ref{Methodology} introduces the ML-FTLE estimator, the Poincar\'{e} style geometric representation, the basis-attractor construction, and the PLSR fusion step. Section~\ref{Results} validates the framework on benchmark transient chaos datasets and analyzes metric-specific robustness. Section~\ref{conclusion} summarizes the main findings and discusses future extensions.

\section{Methodology \label{Methodology}}
A finite-time Lyapunov exponent (FTLEs) quantifies the average rate of exponential stretching or contraction of a trajectory over a finite-time interval. For a $k$-dimensional dynamical system, let $\Phi(t_{j+1},t_j)$ represent the tangent evolution over one sampling interval $\Delta t$. In the QR/Benettin analytical baseline calculation, the $m$-th orthonormal tangent direction $\hat{\mathbf{e}}_j^{(m)}$ is advanced by the tangent dynamics and then re-normalized. The stretching factor is defined as $r_j^{(m)} = \|\Phi(t_{j+1},t_j)\,\hat{\mathbf{e}}_j^{(m)}\|_2,$ where $m=1,2,\ldots,k$. The corresponding finite-time Lyapunov exponent $n$ steps starting at $t_0$ is
\begin{equation}
	\hat{\lambda}_{\mathrm{QR}}^{(m)}(t_0,n)	= \frac{1}{n\Delta t} \sum_{j=0}^{n-1}\ln r_j^{(m)} .
	\label{eq:qr_ftle}
\end{equation}
This normalization by the physical integration time $n\Delta t$ is important when the sampling interval is not unity. Computing Eq.~(\ref{eq:qr_ftle}) requires explicit access to the system equations and their tangent dynamics, which is unavailable for experimental or field acquired scalar time series. The formulation introduced below circumvents this requirement entirely.

\subsection{KNN Forecast Divergence via GMAE (ML-FTLE)}
\label{sec:ml_ftle}
We introduce the ML-FTLE framework as a data-driven and model-free method to estimate the local finite-time Lyapunov exponent directly from a univariate time series. This framework quantifies local chaotic divergence by tracking the temporal growth rate of out-of-sample forecast errors via a machine learning (ML) algorithm based on $k$-nearest-neighbor ($k$-NN) regression. The ML-FTLE block serves as a finite-time, window resolved extension of the machine learning largest Lyapunov exponent estimator, wherein forecast error growth quantified via the geometric mean absolute error (GMAE) acts as a direct proxy for trajectory divergence in chaotic time series.

Given a univariate time series $\{x_i\}_{i=1}^{N}$ sampled at interval $\Delta t$, each analysis window is first standardized. This pre-scaling ensures that all subsequent calculations are dimensionless and invariant to local amplitude drift. We utilize a robust, median-based scaling $\tilde{x}_i = (x_i - \hat{\mu}_{1/2}) / \hat{\sigma}_{\mathrm{rob}}$, where $\hat{\mu}_{1/2}$ represents the sample median. To ensure numerical stability, the scale factor $\hat{\sigma}_{\mathrm{rob}}$ adopts a two-stage rule. It equals the interquartile range ($Q_{0.75} - Q_{0.25}$) if this value meets or exceeds the precision threshold $\varepsilon = 10^{-12}$. Conversely, if the interquartile range collapses below this threshold, the scale factor falls back to $s_x + \varepsilon$, where $s_x$ is the sample standard deviation and $Q_p$ denotes the $p$-th empirical quantile. The interquartile range (IQR) serves as the primary scale estimator due to its robustness against extrema. The standard deviation acts as a fallback only when the temporal window is so nearly constant that the IQR collapses below the numerical floor $\varepsilon$. In both cases, this floor prevents division by zero without biasing the scaled values. For a prediction horizon $h \in [1, H]$, each sample within the scaled window is first embedded into an $m$-dimensional delay-coordinate vector using embedding dimension $m$ and lag $\tau$:
\begin{equation}
    \mathbf{z}_i = \bigl(\tilde{x}_i,\; \tilde{x}_{i+\tau},\;
                         \ldots,\; \tilde{x}_{i+(m-1)\tau}\bigr)
    \in \mathbb{R}^{m}, \nonumber
    \label{eq:delay_embed}
\end{equation}
yielding $N_{\mathrm{vecs}} = W - (m-1)\tau - h$ input target pairs
\begin{equation}
    \mathbf{z}_i \;\longmapsto\; y_{i,h} = \tilde{x}_{i+(m-1)\tau+h},
    \qquad i = 1, \ldots, N_{\mathrm{vecs}}\,. \nonumber
    \label{eq:forecast_pairs}
\end{equation}

By utilizing this delay-coordinate formulation, the algorithm establishes a localized, short-history temporal context for the forecast. Consequently, the $k$-NN neighbor searches are evaluated across multidimensional historical windows that capture the underlying dynamical flow, rather than relying strictly on isolated, instantaneous scalar observations. For each horizon, the estimator thereby learns a local map, $\mathbf{z}_i~\mapsto~\tilde{x}_{i+(m-1)\tau+h}$ and measures how the resulting forecast error grows with $h$. To prevent look-ahead leakage, a chronological train-test split is applied, preserving the first $(1-r_{\mathrm{test}})$ fraction of pairs for training and the remainder for evaluation. Future values are predicted using an inverse-distance weighted $k$-NN
regressor
\begin{equation*}
    \hat{y}_h(\mathbf{z}) =
    \frac{\displaystyle\sum_{j \in \mathcal{N}_k(\mathbf{z})}
          w_j \, y_{j,h}}
         {\displaystyle\sum_{j \in \mathcal{N}_k(\mathbf{z})} w_j}
\end{equation*}
where $\mathcal{N}_k(\mathbf{z})$ denotes the local $k$-point neighborhood of the reference state $\mathbf{z}$ within the reconstructed phase space $\mathbb{R}^m$. The weight $w_j = (\|\mathbf{z} - \mathbf{z}_j\| + \varepsilon)^{-1}$ scales inversely with the Euclidean separation to prioritize structurally proximate states, with $\varepsilon = 10^{-12}$ acting as a regularizer to preclude numerical singularities in the limit of exact spatial recurrence.

In a chaotic regime, microscopic perturbations in the reconstructed state map to exponentially diverging trajectories. To retain the full dynamical spectrum of this separation, the maximal forecast errors remain strictly untruncated. Rather, the out-of-sample absolute errors are aggregated using the Geometric Mean Absolute Error (GMAE):
\begin{equation}
\operatorname{GMAE}(h) = \exp\!\left(\frac{1}{n_{\text{test}}}\sum_{i\in\mathcal{I}_{\text{test}}^{(h)}} \ln\max\!\bigl(|y_{i,h}-\hat{y}_h(\mathbf{z}_i)|,\, \varepsilon\bigr)\right) \nonumber
\end{equation}
where $\mathcal{I}_{\text{test}}^{(h)}$ is the test index set for horizon $h$.

This metric mathematically parallels the temporal averaging formalism of classical Lyapunov analysis by Eq.~(\ref{eq:qr_ftle}), wherein exponential phase space separation is quantified via mean logarithmic growth. For locally hyperbolic dynamics, the empirical trajectory divergence undergoes exponential over finite prediction horizons, $\operatorname{GMAE}(h) \approx A\,e^{\hat{\lambda}_{\mathrm{ML}}\,h\,\Delta t}$. The corresponding logarithmic projection yields a linear temporal scaling relation:
\begin{equation*}
\ln[\operatorname{GMAE}(h)] \approx \hat{\lambda}_{\mathrm{ML}} (h\Delta t) + \ln A.
\end{equation*}

\begin{figure*}
	\includegraphics[width=1\linewidth]{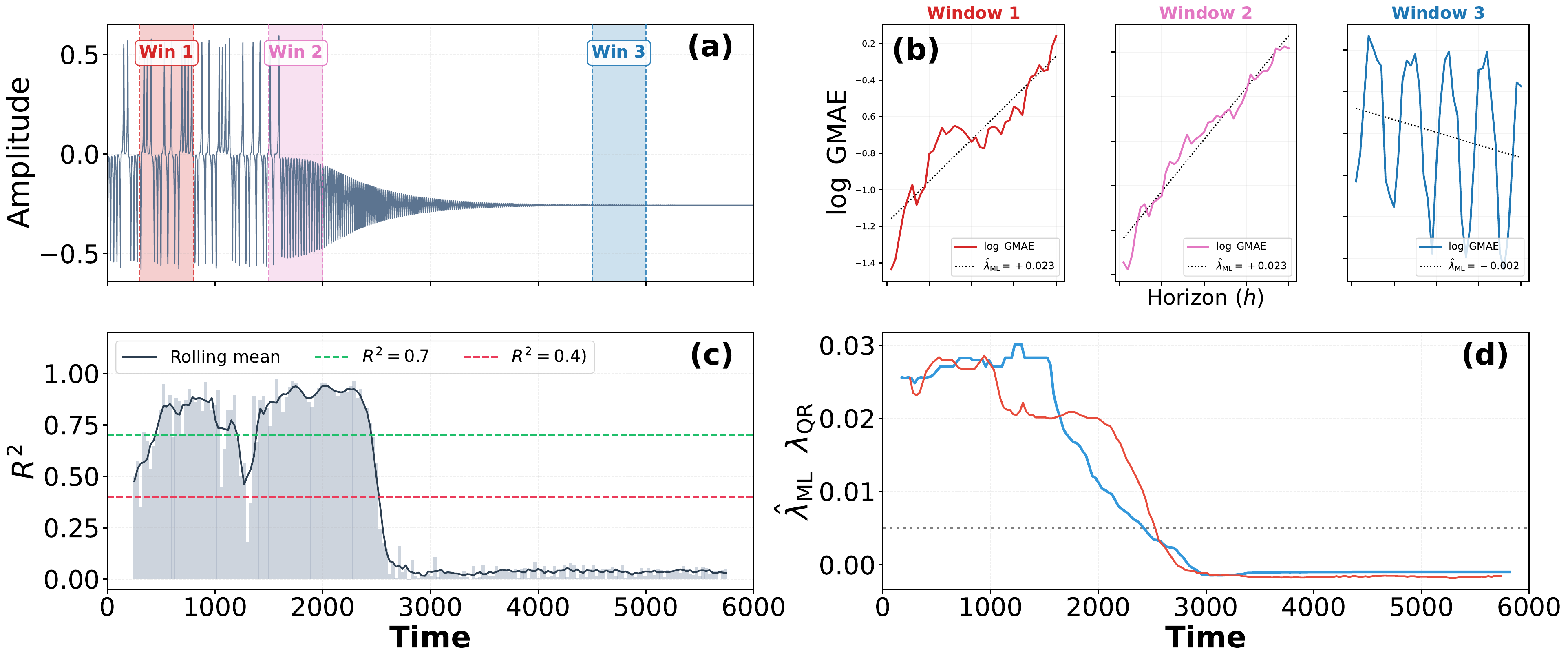}
	\caption{Mechanics of the ML-FTLE estimator. \textbf{(a)} The scalar observable with three representative sliding windows highlighted. \textbf{(b)} The log-transformed geometric mean absolute error versus prediction horizon for the corresponding windows. Windows 1 and 2, which reside in the chaotic regime, exhibit robust linear growth indicative of exponential divergence, while Window 3, located in the stable regime, lacks deterministic growth. \textbf{(c)} Temporal evolution of the linear fit $R^2$ confirming that the exponential divergence model is highly valid ($R^2 > 0.7$) strictly during the chaotic transient. \textbf{(d)} A temporal comparison of the data-driven ML-FTLE ($\hat{\lambda}_{\mathrm{ML}}$) indicated by the red line and the analytical ground truth QR-FTLE  ($\lambda_{\mathrm{QR}}$) baseline represented by the blue line. This demonstrates the ability of the proxy to accurately capture the macroscopic regime shift entirely from scalar data.}
    \label{Fig_1}
\end{figure*}

The estimated Lyapunov scale coefficient ($\hat{\lambda}_{\mathrm{ML}}$) is extracted as the ordinary least squares (OLS) slope of $\ln[\operatorname{GMAE}(h)]$ regressed on $h\Delta t$ over the valid horizon set $\mathcal{L}$, defined as the subset of horizons $h \in \{1,\dots,H\}$ for which $\operatorname{GMAE}(h) \in (0, \infty)$, thereby selectively retaining only those prediction horizons exhibiting finite, positive divergence (i.e., where the out-of-sample error strictly surpasses the numerical precision floor $\varepsilon$).  If $|\mathcal{L}| < 3$, the window is flagged as degenerate, the estimator returns $\hat{\lambda}_{\mathrm{ML}} = 0$ rather than propagating a numerically unreliable slope. For all non-degenerate windows, the Lyapunov proxy is computed as
\begin{equation}
    \hat{\lambda}_{\mathrm{ML}} = \frac{\displaystyle\sum_{h \in \mathcal{L}} \bigl(h\Delta t - \overline{h\Delta t}\bigr) \bigl(\ln[\operatorname{GMAE}(h)] - \overline{\ln[\operatorname{GMAE}(h)]}\bigr)}{\displaystyle\sum_{h \in \mathcal{L}} \bigl(h\Delta t - \overline{h\Delta t}\bigr)^{2}},
    \label{eq:ml-ftle}
\end{equation}
where $\overline{h\Delta t}$ and $\overline{\ln[\operatorname{GMAE}(h)]}$ represent the arithmetic means over $\mathcal{L}$, and the coefficient of determination ($R^2$) of this linear fit is retained as a dynamic reliability metric, assessing the robustness of the local exponential divergence approximation (see Fig.~\ref{Fig_1}).

\subsection{Poincar\'{e}-based Geometric-Guided FTLE Estimator}
\label{sec:poincare_ftle}
We propose a complementary class of FTLE proxies derived exclusively from the structural evolution of lagged Poincar\'{e} return maps, evaluated over sliding temporal windows of the scalar observable. This geometric framework operates across three distinct phases, namely geometric section construction, inter-window structural dissimilarity quantification, and a supervised topological projection calibrated directly against the predictive $\hat{\lambda}_{\mathrm{ML}}$ reference by Eq.~(\ref{eq:ml-ftle}).

For a temporal window of $W$ samples centered at time $t_c$, the lagged Poincar\'{e} return map is constructed from the phase-space coordinates $(x_{i},\, x_{i+\ell})$, where $\ell$ denotes the embedding lag. These coordinates are mapped onto a Boolean occupancy grid, $\mathbf{G} \in \{0,1\}^{B \times B}$, by linearly partitioning the spatial extrema $[v_{\min}, v_{\max}]$ into $B$ uniform bins, utilizing a clipping function to capture boundary intersections. This per-window normalization ensures that the reconstructed trajectory spans the entire $B \times B$ spatial domain independent of the amplitude, making inter-window structural comparisons invariant to amplitude drift. To mitigate discretization artifacts at low spatial resolutions, consecutive state vectors are interpolated via the Bresenham line algorithm\cite{bresenham1965algorithm}, yielding a topologically continuous binary representation. Finally, the empirical occupancy distribution is formalized as the positive normalized histogram $\mathbf{h} = \mathbf{g}/\lVert\mathbf{g}\rVert_1$, where $\mathbf{g}=\operatorname{vec}(\mathbf{G})+\varepsilon$.

To systematically track the structural deformation of the attractor over time, four complementary dissimilarity measures are evaluated between any arbitrary pair of temporal occupancy grids ($\mathbf{G}_p,\mathbf{G}_q$), are defined as follows:
\begin{align}
	D_{\mathrm{JSD}}(p,q) &=
	\left[\tfrac{1}{2}D_{\mathrm{KL}}(\mathbf{h}_p\|\mathbf{m}_{pq})
	+\tfrac{1}{2}D_{\mathrm{KL}}(\mathbf{h}_q\|\mathbf{m}_{pq})\right]^{1/2},\nonumber\\ 
	D_{\mathrm{SSIM}}(p,q) &=
	\tfrac{1}{2}\left[1-\mathrm{SSIM}(\mathbf{G}_p,\mathbf{G}_q)\right],\nonumber\\ 
	D_{\mathrm{HDF}}(p,q) &=
	\max\{d_H(\mathcal{A}_p,\mathcal{A}_q),\,d_H(\mathcal{A}_q,\mathcal{A}_p)\},\nonumber\\ 
	D_{\mathrm{IOU}}(p,q) &=
	1-\frac{|\mathcal{A}_p\cap\mathcal{A}_q|}{|\mathcal{A}_p\cup\mathcal{A}_q|}, \nonumber
\end{align}
where $\mathbf{h}_p$ and $\mathbf{h}_q$ denote the normalized spatial probability distributions of the respective grids and $\mathbf{m}_{pq} = \tfrac{1}{2}(\mathbf{h}_p+\mathbf{h}_q)$ is the pointwise mean probability distribution of the Jensen Shannon Divergence (JSD). Furthermore $d_H(\mathcal{A}_p,\mathcal{A}_q)=\max_{a\in\mathcal{A}_p}\min_{b\in\mathcal{A}_q}\lVert a-b\rVert_2$ defines the directed Hausdorff distance (HDF) between occupied coordinate sets where $\mathcal{A}_p$ and $\mathcal{A}_q$ represent the active cell index sets of $\mathbf{G}_p$ and $\mathbf{G}_q$ respectively. Lastly, SSIM and IOU designate the Structural Similarity Index and Intersection over Union~\cite{Wang2004}. Together, these metrics capture the distributional, structural, geometric, and overlap-based dimensions of phase-space deformation, which collectively serve as morphological indicators for shifts in the underlying Lyapunov spectrum.

To mitigate the computational redundancy of evaluating the entire temporal sequence, we construct a compact basis of $N_{\mathrm{rep}}$ characteristic structural anchors spanning the $N_w$ temporal windows using a greedy max-min diversity algorithm. Initialized at the median geometric state of the trajectory, the algorithm iteratively extracts each new representative $r^*$ by maximizing its minimum dissimilarity to the existing basis set $\mathcal{S}$:
\begin{equation}
r^* = \operatorname*{arg\,max}_{i \notin \mathcal{S}} \min_{s \in \mathcal{S}} D^{(m)}(i,\,s), \nonumber  
\label{eq:greedy}
\end{equation}
this selection is subject to a minimum temporal separation constraint of $|i - s| \ge \lceil N_w\,\delta_{\min}\rceil$ for all $s \in \mathcal{S}$. For each metric $m$, the normalized structural proximity between a temporal window $i$ and a basis state $r \in \mathcal{S}$ is defined as
\begin{align}
	D_{\max}^{(m)} &= \max_{1\le p<q\le N_w} D^{(m)}(p,q), \nonumber\\
	c_{i,r}^{(m)} &= \max\!\left(0,\,
	1 - \frac{D^{(m)}(i,\,r)}{D_{\max}^{(m)}+\varepsilon}\right).
	\label{eq:closeness scores}
\end{align}
This formulation projects the empirical dissimilarity to a $[0,1]$ interval relative to the maximum geometric separation for the respective metric. The evolving system state is thus embedded within a closeness matrix $\mathbf{C}^{(m)}\in\mathbb{R}^{N_w\times N_{\mathrm{rep}}}$, mapping each temporal window onto the $N_{\mathrm{rep}}$ basis attractor. The columns of $\mathbf{C}^{(m)}$ are non-orthogonal, as distinct basis states may exhibit structurally collinear responses to the same geometric deformation of the reconstructed attractor. To map this topological representation, we utilize a one-component PLSR model to extract a supervised latent component from $\mathbf{C}^{(m)}$ that maximally covaries with the ML-FTLE ($\hat{\lambda}_{\mathrm{ML}}$) sequence, yielding a robust geometry to instability calibration while bypassing the ill-conditioned direct regression on structurally redundant predictors \cite{Abdi2010Partial,Qin2020Bridging}.

In this framework, $\mathbf{X} \in \mathbb{R}^{N_w \times N_{\mathrm{rep}}}$ represent the standardized closeness matrix aligned in time and $\mathbf{y} \in \mathbb{R}^{N_w}$ as the mean-centered ML-FTLE sequence. A one-component PLSR model isolates the optimal projection weight vector $\mathbf{w}~=~\mathbf{X}^\top\mathbf{y}/\lVert \mathbf{X}^\top\mathbf{y}\rVert_2$. The sequence is projected onto this vector to extract the latent geometric component $\mathbf{t} = \mathbf{X}\mathbf{w}$, which is calibrated back to the Lyapunov scale via the ordinary least squares regression coefficient $q = (\mathbf{t}^\top \mathbf{y})/(\mathbf{t}^\top \mathbf{t})$, yielding the geometric-guided FTLE proxy $\hat{\lambda}_{\mathrm{geo}} = \mathbf{t}q$. While the equation-free ML-FTLE acts as the calibration target for PLSR, the analytical QR-FTLE ($\lambda_{\mathrm{QR}}$) reference by Eq.~(\ref{eq:qr_ftle}) is reserved strictly for independent out-of-sample validation within the synthetic benchmarks.

\begin{algorithm}[H]
    \caption{ML-FTLE $\And$ Poincar\'{e}-based geometric-guided FTLE}\label{alg}
    \begin{algorithmic}[1]
        \Statex \textbf{Phase 1: ML-FTLE Estimation}
        \For{each temporal window $w \subset \{x_i\}$ with center $t_c$}
            \State $\tilde{w} \leftarrow \textsc{Robust Scale}(w)$
            \State Compute $\operatorname{GMAE}(h)$ via out-of-sample $k$-NN for $h \in [1, H_{\max}]$
            \State $\mathcal{L} \leftarrow \{h \mid \operatorname{GMAE}(h) > 0\}$
            \State $\hat{\lambda}_c \leftarrow \text{OLS slope of } \ln[\operatorname{GMAE}(h)] \text{ vs } (h \Delta t) \quad \forall h \in \mathcal{L}$
            \State Append $t_c \to \mathbf{T}_{\mathrm{ML}}$, $\hat{\lambda}_c \to \hat{\boldsymbol{\lambda}}_{\mathrm{ML}}$ (refer Eq.~(\ref{eq:ml-ftle}))
        \EndFor
        
        \Statex % Adds a visual line break
        \Statex \textbf{Phase 2: Geometric Extraction \& Basis Selection}
        \For{each temporal window $w \subset \{x_i\}$ with center $t_c$}
            \State $\mathbf{G}_w, \mathbf{h}_w \leftarrow \textsc{Poincar\'{e} Grid}(w, \text{lag}=\ell, \text{bins}=B)$
            \State Append $\mathbf{G}_w \to \mathcal{G}$
        \EndFor

        \State \textbf{Initialize} the representative set $\mathcal{S} \leftarrow \left\{ \lfloor |\mathcal{G}|/2 \rfloor \right\}$ 
        \While{$|\mathcal{S}| < N_{\mathrm{rep}}$}
            \State $r^* \leftarrow \operatorname*{arg\,max}_{i \notin \mathcal{S}} \min_{s \in \mathcal{S}} D^{(m)}(i, s)$ 
            \Statex \qquad \text{subject to } $|i - s| \ge \lceil N_w\,\delta_{\min}\rceil \quad \forall s \in \mathcal{S}$
            \State $\mathcal{S} \leftarrow \mathcal{S} \cup \{r^*\}$
        \EndWhile

        \State $D_{\max}^{(m)} \leftarrow \max_{p<q}D^{(m)}(p,q)$
        \For{each window $i \in [1, N_w]$ and representative $r \in \mathcal{S}$}
            \State $c_{i,r}^{(m)} \leftarrow \max\!\left(0, 1 - \frac{D^{(m)}(i,r)}{D_{\max}^{(m)}+\varepsilon}\right)$ (refer Eq.~(\ref{eq:closeness scores}))
        \EndFor

        \Statex % Adds a visual line break
        \Statex \textbf{Phase 3: Geometry-Guided Projection (PLSR)}
        \State $\mathbf{C}^{(m)}_{\parallel} \leftarrow \textsc{Interpolate}(\mathbf{C}^{(m)}, \text{onto } \mathbf{T}_{\mathrm{ML}})$
        \State $\mathbf{X} \leftarrow \textsc{Standardize}(\mathbf{C}^{(m)}_{\parallel})$
        \State $\mathbf{y} \leftarrow \textsc{Mean Center}(\hat{\boldsymbol{\lambda}}_{\mathrm{ML}})$

        \State $\mathbf{w} \leftarrow \frac{\mathbf{X}^\top \mathbf{y}}{\lVert \mathbf{X}^\top \mathbf{y} \rVert_2}$ 
        \State $\mathbf{t} \leftarrow \mathbf{X} \mathbf{w}$ \Comment{Extract latent geometric component}
        \State $q \leftarrow \frac{\mathbf{t}^\top \mathbf{y}}{\mathbf{t}^\top \mathbf{t}}$ \Comment{Compute OLS}

        \State $\hat{\lambda}_{\mathrm{geo}} \leftarrow \mathbf{t} q$ \Comment{Compute geometric-guided FTLE}

        \State \Return $\hat{\boldsymbol{\lambda}}_{\mathrm{ML}}, \mathbf{t}, \hat{\lambda}_{\mathrm{geo}}$
    \end{algorithmic}
\end{algorithm}

\subsection{Data Generation and System Dynamics}\label{data_generation}
To rigorously evaluate the performance of the proposed estimators, we generated synthetic time-series datasets ($\mathcal{D}_k$) and their corresponding analytical ground-truth FTLE ($\lambda_{\mathrm{QR}}$) using a nonlinear memcapacitive oscillator~\cite{wang2020multistability}. This system was explicitly selected for its complex dynamical repertoire, enabling the precise isolation of severe geometric phase space transitions. To comprehensively test the algorithms, three distinct transient chaos scenarios were simulated to capture varying rates of structural decay, such as a gradual damping from chaos to a fixed point ($\mathcal{D}_1$), an abrupt chaotic collapse to a fixed point ($\mathcal{D}_2$), and a transition from chaos to a periodic state ($\mathcal{D}_3$). Because these transient phenomena feature varying rates of structural decay, making them an ideal benchmark system to validate how accurately the ML-FTLE and Geometry-guided FTLE frameworks map macroscopic geometric collapses onto the Lyapunov scale.

\section{Results \label{Results}}
\subsection{Evaluation of the Predictive ML-FTLE Framework}
%%%%%%%%%%%%%%%%%%%%%%%%%%%%%%%%%%%%%%%%%%%%%%%%%%%%%%%%%
%%--------------Chaos to Fixed Point---------------
%%%%%%%%%%%%%%%%%%%%%%%%%%%%%%%%%%%%%%%%%%%%%%%%%%%%%%%%%%
To systematically validate the efficacy of the proposed model-free algorithm, the ML-FTLE ($\hat{\lambda}_{\mathrm{ML}}$) was evaluated against the exact analytical QR-FTLE ($\lambda_{\mathrm{QR}}$) ground truth across a highly non-stationary dynamical transition. The selected synthetic dataset ($\mathcal{D}_1$) exhibits a prolonged regime of transient chaos that gradually damps into a stable steady state, presenting a rigorous test of the estimator's capacity to track geometric phase-space collapses strictly from a scalar observable. 

The macroscopic dynamics and the internal mechanics of the ML-FTLE estimator are detailed in Fig.~\ref{Fig_1}, where Fig.~\ref{Fig_1}(a) illustrates the raw scalar time series. To evaluate local divergence characteristics, out-of-sample forecast errors were computed across sliding temporal windows. As predicted by the theoretical framework and demonstrated in Fig.~\ref{Fig_1}(b), the logarithmic Geometric Mean Absolute Error (log-GMAE) exhibits strict linear growth against the prediction horizon during the chaotic regime (Windows 1 and 2). This linear expansion confirms the presence of deterministic exponential trajectory divergence, thereby validating the local hyperbolic assumption. Conversely, within the stable regime (Window 3), trajectory separation ceases, and the out-of-sample forecast errors become dominated by stochastic noise rather than deterministic stretching, driving a complete breakdown of the linear growth model.

The statistical robustness of this exponential divergence model is tracked continuously in Fig.~\ref{Fig_1}(c) via the coefficient of determination ($R^2$) of the ordinary least squares fit. During the transient chaotic phase, the $R^2$ values remain consistently high ($R^2 > 0.75$), statistically validating the local Lyapunov proxy extraction. At the precise moment the strange attractor collapses, the $R^2$ metric abruptly drops to near zero, confirming that the algorithm inherently registers the transition to a non-chaotic state. Finally, Fig.~\ref{Fig_1}(d) provides a direct temporal comparison between the purely data-driven ML-FTLE ($\hat{\lambda}_{\mathrm{ML}}$) represented by the red line and the equation-driven analytical reference ($\lambda_{\mathrm{QR}}$) indicated by the blue line. 

Despite operating entirely on the scalar observable without prior knowledge of the governing differential equations, the ML-FTLE ($\hat{\lambda}_{\mathrm{ML}}$) successfully captures the macroscopic regime shift, it maintains a strictly positive divergence rate throughout the chaotic transient and accurately mirrors the analytical reference's convergence to a non-positive value as the system settles into a steady state. This strong alignment between $\hat{\lambda}_{\mathrm{ML}}$ and $\lambda_{\mathrm{QR}}$ provides robust empirical validation that GMAE-based forecast divergence slope estimation provides a reliable, equation-free surrogate for the maximum finite-time Lyapunov exponent.

\begin{figure}
	\includegraphics[width=1\linewidth]{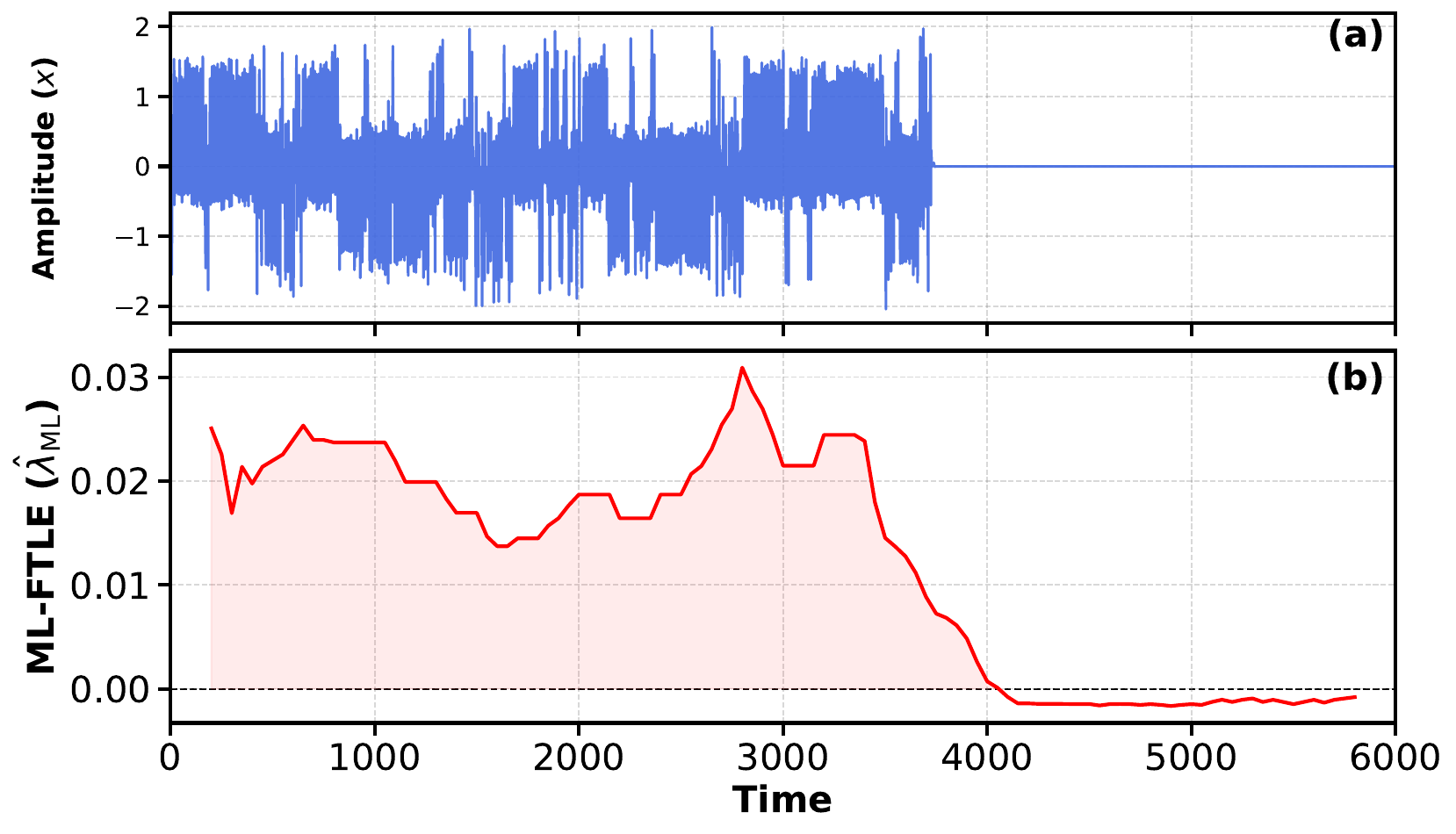}
    \caption{\textbf{(a)} The scalar observable undergoing an abrupt topological collapse from transient chaos into a stationary state near $t \approx 3800$. \textbf{(b)} Temporal evolution of the predictive ML-FTLE $\hat{\lambda}_{\mathrm{ML}}$. The estimator distinguishes the macroscopic regime shift directly from the scalar data, maintaining a positive Lyapunov exponent during the chaotic transient before dropping sharply as the system stabilizes.}
    \label{Fig_2}
\end{figure}

%%%%%%%%%%%%%%%%%%%%%%%%%%%%%%%%%%%%%%%%%%%%%%%%%%%%%%%%%%%%%
%%--------Chaos to fixed point (abruptly)-------------
%%%%%%%%%%%%%%%%%%%%%%%%%%%%%%%%%%%%%%%%%%%%%%%%%%%%%%%%%%%%%%
To further demonstrate the generalizability of the ML-FTLE ($\hat{\lambda}_{\mathrm{ML}}$) estimator across distinct classes of dynamical regime shifts, the algorithm~(\ref{alg}) was evaluated on datasets $\mathcal{D}_2$ and $\mathcal{D}_3$. Figure~\ref{Fig_2} details an abrupt dynamical collapse in $\mathcal{D}_2$, a class of sudden topological transitions that typically degrade the tracking accuracy of standard windowed, data-driven predictive algorithms due to inherent temporal lags. Figure~\ref{Fig_2}(a) illustrates the signal (scalar observable) variance suddenly collapsing at $t \approx 3800$, signifying the destruction of the chaotic attractor. Despite this abrupt discontinuity, the $\hat{\lambda}_{\mathrm{ML}}$ depicted in Fig.~\ref{Fig_2}(b) tracks the structural regime transition with high precision. The estimated instability scale drops sharply, crossing the predefined critical threshold ($\lambda_{\mathrm{thresh}} = 0.005$) into the stable regime in direct correspondence with the sudden collapse of the chaotic attractor.

\begin{figure}
	\includegraphics[width=1\linewidth]{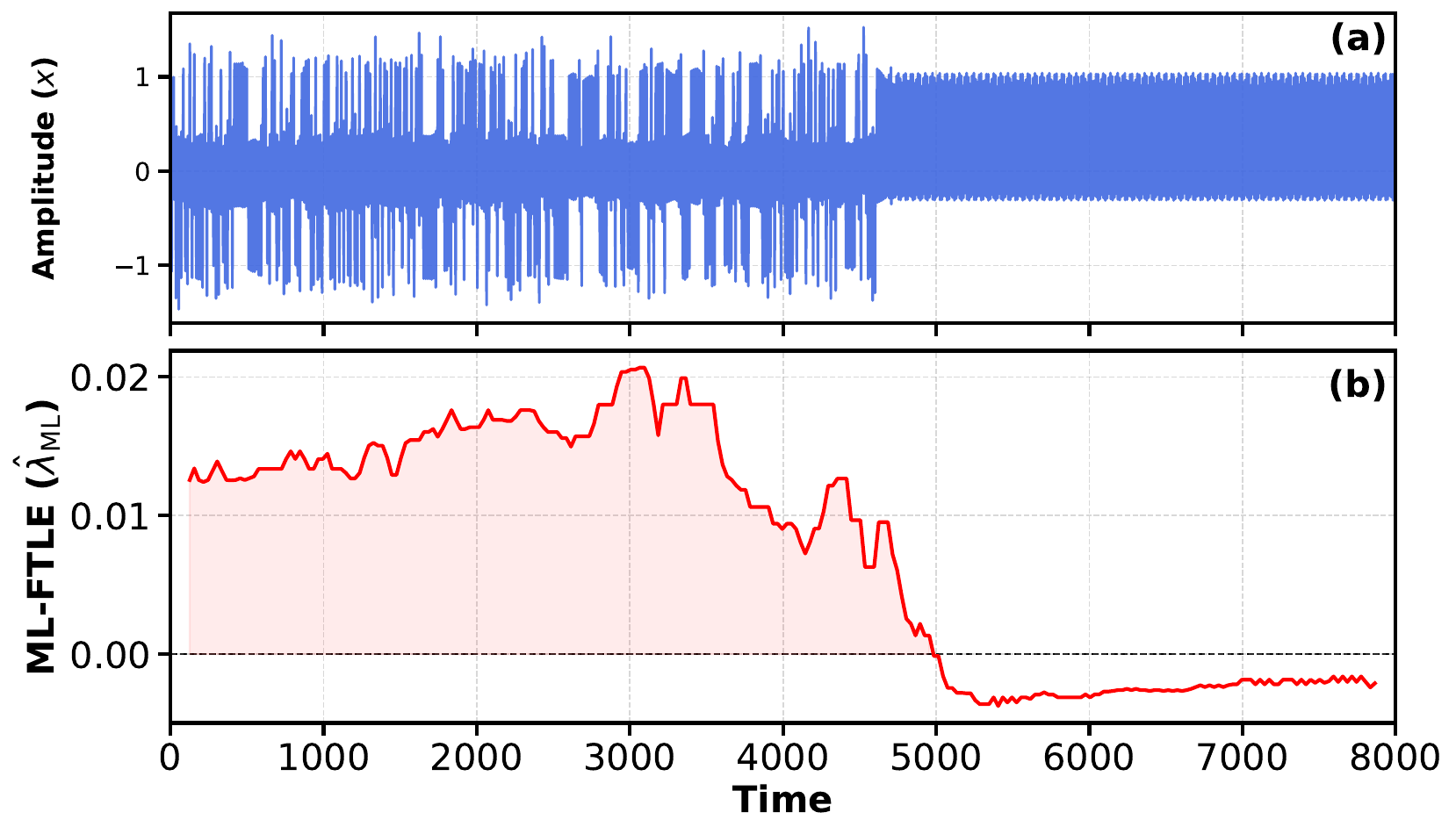}
    \caption{\textbf{(a)} The scalar observable transitioning from transient chaotic oscillations into an asymptotic periodic regime. \textbf{(b)} Temporal evolution of the predictive ML-FTLE proxy $\hat{\lambda}_{\mathrm{ML}}$ derived directly from the scalar data. The estimator accurately isolates the macroscopic transition, dropping sharply to cross the stability threshold as the chaotic attractor dissipates into the periodic orbit.}
    \label{Fig_3}
\end{figure}

%%%%%%%%%%%%%%%%%%%%%%%%%%%%%%%%%%%%%%%%%%%%%%%%%%%%%%
%%---------Chaos to Periodic State-------------
%%%%%%%%%%%%%%%%%%%%%%%%%%%%%%%%%%%%%%%%%%%%%%%%%%%%%%
Figure~\ref{Fig_3} illustrates the estimator's performance on dataset $\mathcal{D}_3$. The raw scalar observable displayed in Fig.~\ref{Fig_3}(a) transitions from transient chaos into a stable high-frequency periodic state near $t \approx 4700$. As shown in Fig.~\ref{Fig_3}(b), the ML-FTLE ($\hat{\lambda}_{\mathrm{ML}}$) reliably identifies the chaotic regime by maintaining a strongly positive divergence rate. Upon the macroscopic regime shift, the proxy exhibits a sharp decline, crossing the stability threshold. This confirms the estimator's capacity to accurately classify the system's convergence into a non-chaotic periodic orbit, where the slight temporal lag is a standard predictable consequence of the sliding window forecast construction, which is  rectified upon integrating the model into the Geometry-guided FTLE framework (Sec.~\ref{geometric_FTLE}). 
Collectively, the results from $\mathcal{D}_1$, $\mathcal{D}_2$, and $\mathcal{D}_3$ confirm that the entirely data-driven ML-FTLE framework can reliably map complex, highly non-stationary scalar transitions to the quantitative Lyapunov scale. It successfully distinguishes transient chaos from both stable equilibria and periodic orbits, functioning entirely without prior knowledge of the governing differential equations.

\subsection{Performance of the Geometric-Guided FTLE Estimator \label{geometric_FTLE}}
\begin{figure}
	\centering
	\includegraphics[width=1\linewidth]{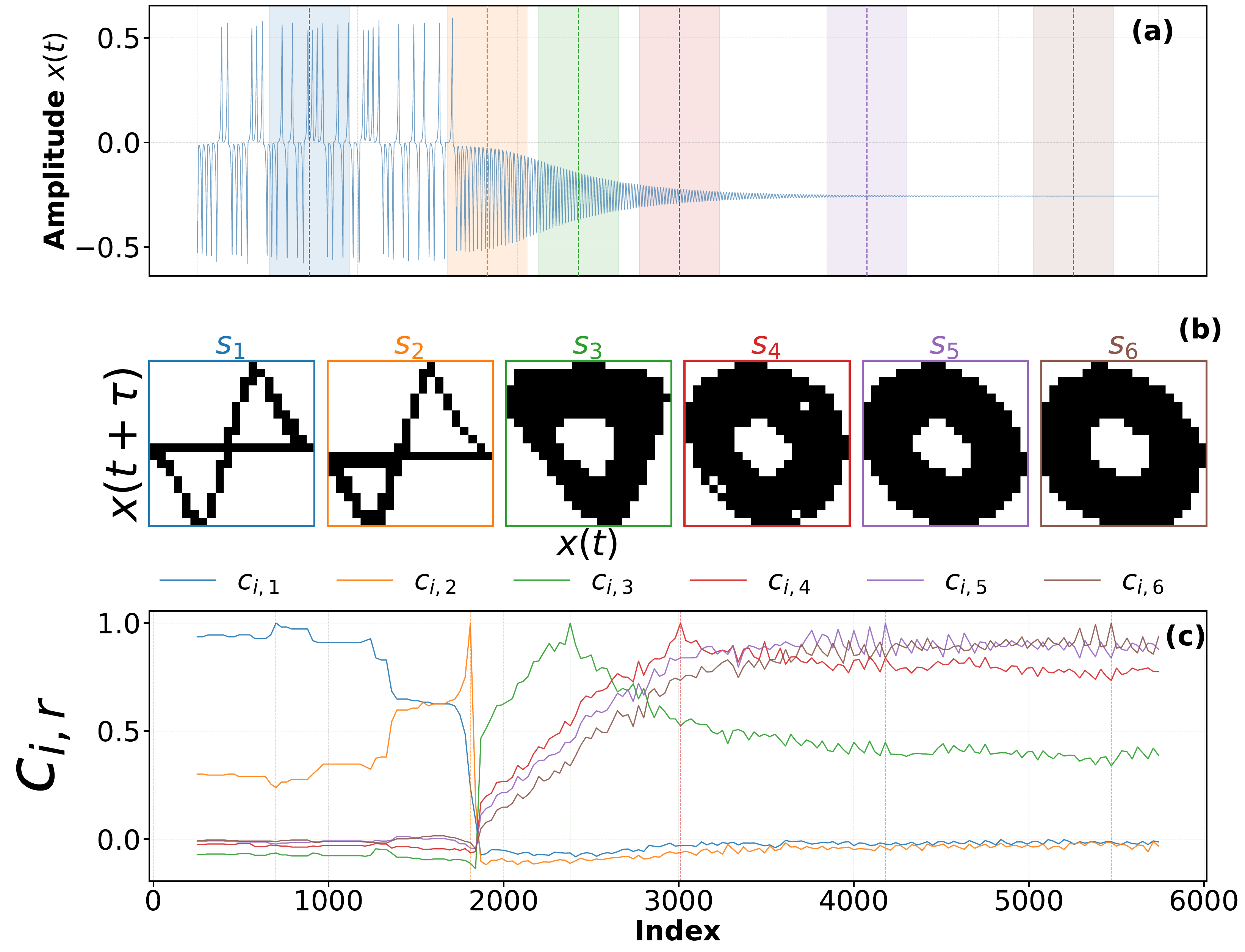}
	\caption{Visualization of the geometry guided ML-FTLE decomposition process. \textbf{(a)} The raw scalar observable plotted in blue. Colored vertical bands indicate the temporal representatives ($\mathcal{S}$) extracted by the selection algorithm. \textbf{(b)} The 2D Poincar\'{e} occupancy grids mapping ($x(t)$ - $x(t+\tau)$) corresponding to the selected temporal anchors. These binary grids constitute the structural basis set $\mathcal{S}$ capturing the macroscopic attractor geometry. \textbf{(c)} The continuous temporal evolution of the normalized closeness scores $c_{i,r}$ relative to each basis attractor computed utilizing the SSIM measure.}
	\label{Fig_4}
\end{figure}

%%%%%%%%%%%%%%%%%%%%%%%%%%%%%%%%%%%%%%%%
%------Topological Decomposition process
%%%%%%%%%%%%%%%%%%%%%%%%%%%%%%%%%%%%%%%%
To illustrate the internal mechanics and efficacy of the geometric-guided FTLE framework, we apply the complete analytical pipeline to a non-stationary trajectory exhibiting a distinct dynamical transition. Detailed in Fig.~\ref{Fig_4}, the results utilizing the Structural Similarity Index (SSIM) as the governing topological metric demonstrate how macroscopic geometric shifts are automatically extracted and translated into a continuous Lyapunov proxy. For this specific dataset ($\mathcal{D}_1$), the raw scalar observable was shown in Fig.~\ref{Fig_4}(a), undergoes a clear macroscopic topological transition near the sample index $1800$, shifting from a chaotic oscillatory state into a damped steady state regime. Rather than relying on localized nearest-neighbor tracking, the framework captures this exact dynamical collapse purely through spatial morphology. Subject to a minimum temporal separation constraint ($\delta_{\min}$), the greedy max-min selection algorithm populates the basis set $\mathcal{S}$ by extracting $N_{\mathrm{rep}} = 6$ maximally dissimilar representative windows (highlighted by colored vertical bands), where $N_{\mathrm{rep}}$ acts as a user-defined hyperparameter. 

Figure~\ref{Fig_4}(b) illustrates the underlying 2D Poincar\'{e} occupancy grids corresponding to these temporal reference states, the maximally dissimilar windows naturally partition the trajectory into its two fundamental dynamical regimes. Representatives $\mathcal{S}_1$ and $\mathcal{S}_2$ originate from the chaotic transient and exhibit broadly distributed, complex phase-space topologies. Conversely, Representatives $\mathcal{S}_3$ to $\mathcal{S}_6$ are sampled from the asymptotically stable regime, which is characterized by minimal, highly localized structures indicative of non-chaotic dynamics. This intuitive alignment emerges organically from applying the geometric dissimilarity criterion to a trajectory governed by a single regime shift, rather than being explicitly prescribed by the algorithm. 

Figure~\ref{Fig_4}(c) tracks the continuous temporal evolution of the normalized geometric closeness scores $c_{i,r}$ by Eq.~(\ref{eq:closeness scores}) relative to each of the six basis attractors, which collectively span the similarity matrix $\mathbf{C}$. The structural transition is clearly delineated by these evolving trajectories, proximity to $\mathcal{S}_1$ characterizes the chaotic phase, whereas $\mathcal{S}_2$ acts as a transient structural bridge precisely at the critical transition point. Immediately following the topological collapse, proximity to the localized morphologies ($\mathcal{S}_3$ to $\mathcal{S}_6$) sharply increases and maintains prevalence throughout the stable regime.

\begin{figure}
	\centering
	\includegraphics[width=1\linewidth]{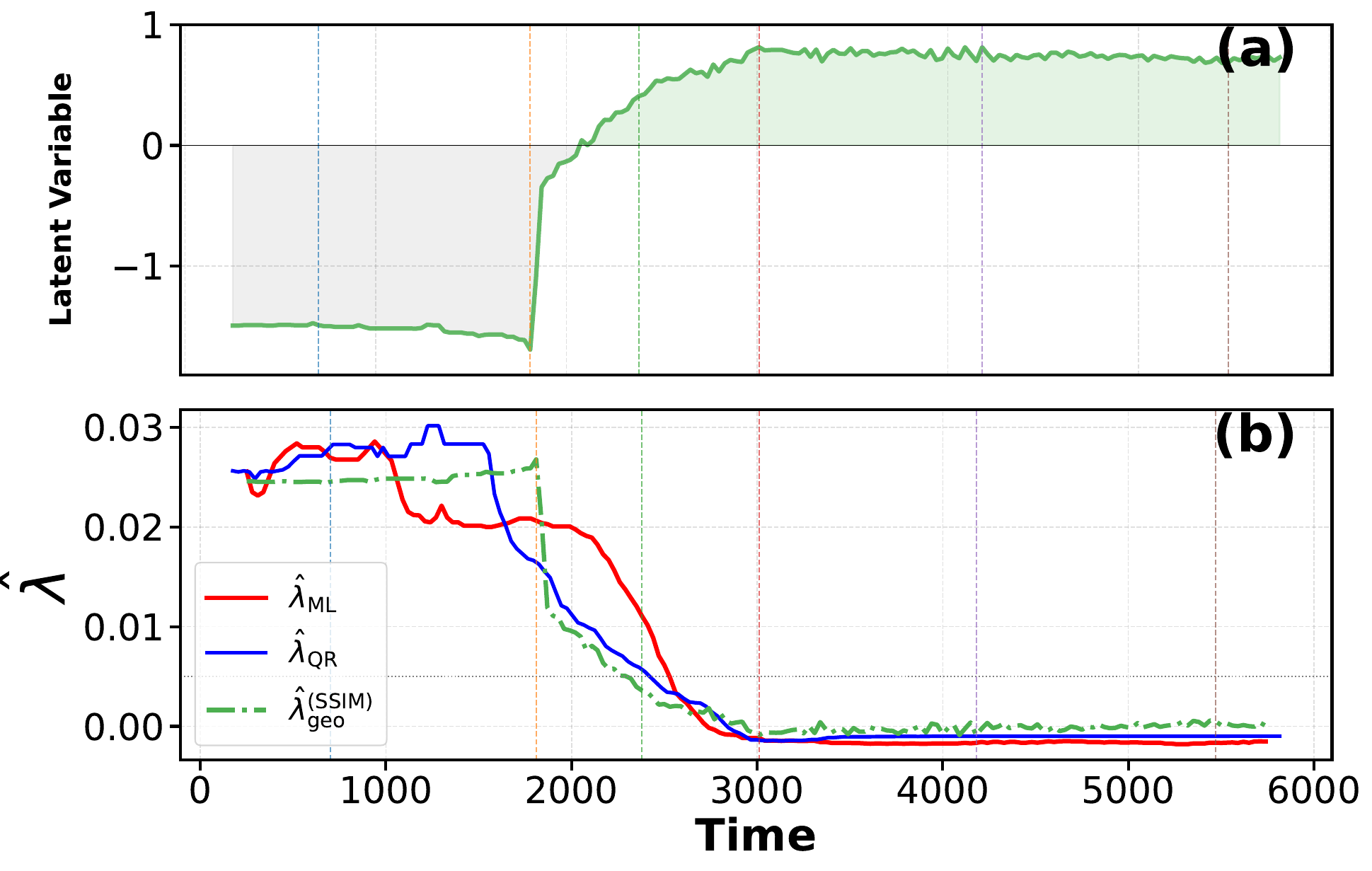}
	\caption{Evaluation of the SSIM derived Geometry-guided FTLE estimator. \textbf{(a)} The latent variable isolates the structural transition from a chaotic regime shaded in grey to a stable state shaded in green. Dashed lines mark the temporal reference states $\mathcal{S}$. \textbf{(b)} The Geometry-guided FTLE $\hat{\lambda}_{\mathrm{geo}}^{(\mathrm{SSIM})}$ indicated by the green line overcomes the temporal lag of the predictive ML-FTLE $\hat{\lambda}_{\mathrm{ML}}$ shown in red. Validated against the analytical QR-FTLE ($\lambda_{\mathrm{QR}}$) ground truth plotted in blue the geometric estimator precisely resolves the abrupt phase space collapse without delay.}
	\label{Fig_5}
\end{figure}

By applying Partial Least Squares Regression, this high-dimensional, purely geometric evolution is projected onto a single latent component calibrated directly to the reference Lyapunov scale $\hat{\lambda}_{\mathrm{ML}}$. The temporal evolution of this latent variable is depicted in Fig.~\ref{Fig_5}(a). The resulting Geometry-guided FTLE ($\hat{\lambda}_{\mathrm{geo}}^{(\mathrm{SSIM})}$) indicated by the green line in Fig.~\ref{Fig_5}(b) smoothly tracks the onset of the dynamical transition and accurately converges during the stable regime. Crucially, it closely mirrors the analytical QR-FTLE ($\hat{\lambda}_{\mathrm{QR}}$) shown as the blue line, achieving a strong Spearman rank correlation of $\rho = 0.94$. This represents a marked improvement over the predictive ML-FTLE ($\hat{\lambda}_{\mathrm{ML}}$) represented by the red line, which yields $\rho = 0.81$ (as further detailed in Sec.~\ref{Quantitative Performance}). This confirms that low resolution binary occupancy grids retain sufficient deterministic structure to track the FTLEs instantaneously, thereby resolving the temporal detection lag observed in the ML-FTLE model.

%%%%%%%%%%%%%%%%%%%%%%%%%%%%%%%%%%%%%%%%%%%%%%%%%%%%%%%%%%%%%%%%%%%%%%
%-------- FTLE comparative analysis (Spearman rank and MCC)
%%%%%%%%%%%%%%%%%%%%%%%%%%%%%%%%%%%%%%%%%%%%%%%%%%%%%%%%%%%%%%%%%%%%%%
\subsection{Quantitative Performance \label{Quantitative Performance}}
\begin{figure*}
    \includegraphics[width=1\linewidth]{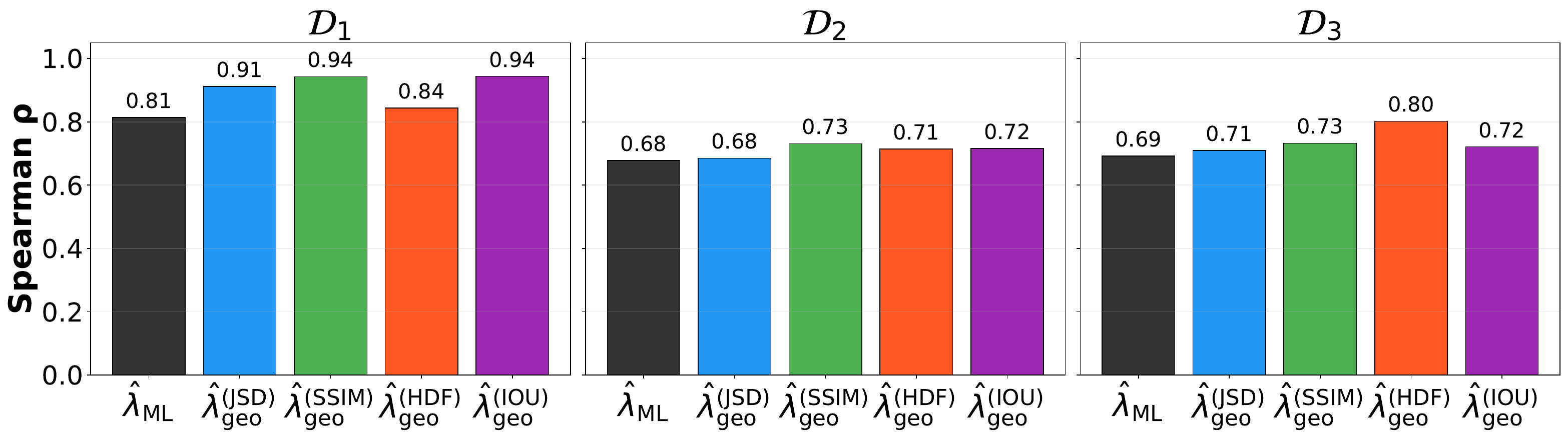}
    \caption{Spearman rank correlation ($\rho$) evaluating continuous tracking of the analytical QR-FTLE ($\lambda_{\mathrm{QR}}$) ground truth. The predictive ML-FTLE ($\lambda_{\mathrm{ML}}$) and four geometry-guided estimators ($\hat{\lambda}_{\mathrm{geo}}^{(m)}$), namely $m \in \{\mathrm{JSD},\,\mathrm{SSIM},\,\mathrm{HDF},\,\mathrm{IOU}\}$. These $\hat{\lambda}_{\mathrm{geo}}^{(m)}$ consistently outperform the baseline during severe dynamical shifts. $\hat{\lambda}_{\mathrm{geo}}^{(SSIM)}$ and $\hat{\lambda}_{\mathrm{geo}}^{(IOU)}$ optimally resolves gradual damping in the dataset $\mathcal{D}_1$ and $\mathcal{D}_2$ while $\hat{\lambda}_{\mathrm{geo}}^{(HDF)}$ demonstrating extreme resilience during abrupt topological transitions in dataset $\mathcal{D}_3$.}
    \label{metric_spearman_rho_vs_qr}
\end{figure*}

\begin{figure*}
    \includegraphics[width=1\linewidth]{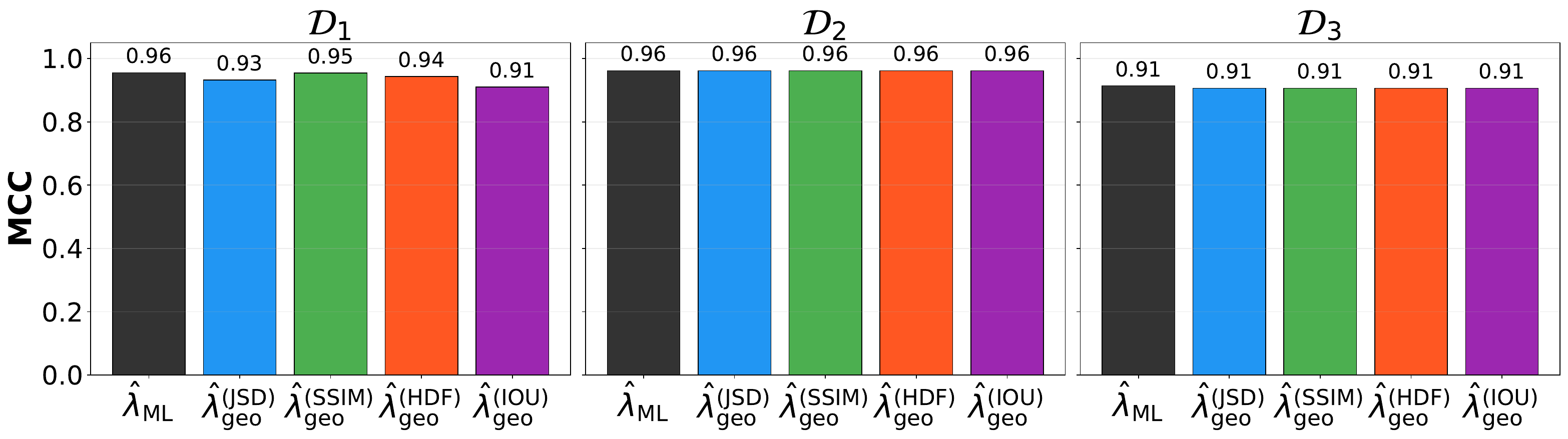}
    \caption{The Matthews Correlation Coefficient (MCC) evaluating the binary classification of the chaotic to stable and periodic regime shift. All methods exhibit exceptional threshold crossing accuracy with $\mathrm{MCC} > 0.90$ demonstrating robust performance convergence during the abrupt dynamical transition.}
    \label{metric_mcc_vs_qr}
\end{figure*}

To rigorously evaluate the fidelity of each method, we utilize the comparative analysis of the two most diagnostically informative metrics, namely the Spearman rank correlation coefficient ($\rho$) and Matthews Correlation Coefficient (MCC). The Spearman $\rho$ quantifies the monotonic agreement between each method and the QR-FTLE ($\hat{\lambda}_{\mathrm{QR}}$) ground truth across the full dynamic range, while the MCC provides a balanced classification score for the binary chaotic/non-chaotic regime distinction, accounting for class imbalance in the threshold-crossing labels. Both metrics are bounded in $[-1,\,1]$, with unity indicating perfect correspondence and zero indicating chance-level agreement. Figures~\ref{metric_spearman_rho_vs_qr} and \ref{metric_mcc_vs_qr} present per-dataset bar charts evaluating two distinct method families namely the predictive ML-FTLE ($\hat{\lambda}_{\mathrm{ML}}$) and the calibrated Geometry-guided FTLE ($\hat{\lambda}_{\mathrm{geo}}^{(m)}$) across four dissimilarity metrics, $m \in \{\mathrm{JSD},\,\mathrm{SSIM},\,\mathrm{HDF},\,\mathrm{IOU}\}$. All estimators are evaluated directly against the analytical $\lambda_{\mathrm{QR}}$ ground truth.

%----------------------------------------------------
%%%%%%%%%%%%%% Spearman rank correlation %%%%%%%%%%%%%
%----------------------------------------------------

To rigorously quantify the continuous temporal alignment between the data-driven proxies and the exact analytical reference, the Spearman rank correlation is defined for each dataset $\mathcal{D}_k$ as:
\begin{equation}
\rho^{(m)}_k = 1 - \frac{6 \displaystyle\sum_{t=1}^{n} d_{t}^{2}}{n\left(n^{2}-1\right)},
\label{eq:spearman}
\end{equation}
where $d_t = \mathrm{rank}\!\left(\lambda_{\mathrm{QR}}(t)\right) - \mathrm{rank}\!\left(\hat{\lambda}^{(m)}(t)\right)$ is the rank differential at temporal window $t$, $\hat{\lambda}^{(m)}$ denotes the proxies for $\hat{\lambda}_{\mathrm{ML}}$ and $\hat{\lambda}_{\mathrm{geo}}$, and $n$ represents the total number of samples in the aligned time series. It is fundamental to underscore that the topological Poincar\'{e}-based geometric-guided FTLE $\hat{\lambda}_{\mathrm{geo}}^{(m)}$ is calibrated strictly against the empirical ML-FTLE ($\hat{\lambda}_{\mathrm{ML}}$) during the PLSR training phase. The analytical QR-FTLE ($\lambda_{\mathrm{QR}}$) remains completely isolated from the fitting procedure and is utilized solely as an independent, out-of-sample ground truth for evaluating $\rho^{(m)}_k$. While Eq.~(\ref{eq:spearman}) presents the conventional formulation for notational simplicity, the computational implementation calculates the Pearson correlation of the rank transformed sequences to robustly account for tied ranks.

The continuous tracking performance, quantified by the Spearman rank correlation ($\rho$) depicted in Fig.~\ref{metric_spearman_rho_vs_qr}, reveals the distinct advantage of the geometric projection $(\hat{\lambda}_{\mathrm{geo}}^{(m)}$). The predictive ML-FTLE ($\hat{\lambda}_{\mathrm{ML}}$) struggles to smoothly track the monotonic evolution of the underlying Lyapunov spectrum, particularly during the drastic geometric shifts of $\mathcal{D}_2$ ($\rho = 0.68$) and $\mathcal{D}_3$ ($\rho = 0.69$). By contrast, projecting the dynamics through the low-dimensional structural basis set systematically improves temporal tracking. Crucially, the optimal dissimilarity metric depends strictly on the physical nature of the transient phase. For $\mathcal{D}_1$, which represents a gradual damping transition, the geometry-guided SSIM ($\hat{\lambda}_{\mathrm{geo}}^{(SSIM)}$) and IOU ($\hat{\lambda}_{\mathrm{geo}}^{(IOU)}$) emerge as the optimal proxies ($\rho = 0.94$). These metrics are highly sensitive to smooth, continuous variations in spatial probability density as the attractor slowly contracts. In the case of an abrupt transition ($\mathcal{D}_2$), $\hat{\lambda}_{\mathrm{geo}}^{(SSIM)}$ maintains a slight analytical edge ($\rho = 0.73$). However, during the severe topological transition of $\mathcal{D}_3$, from a chaotic into a periodic state,  geometry-guided HDF ($\hat{\lambda}_{\mathrm{geo}}^{(HDF)}$) proves to be the most robust method ($\rho = 0.80$). Because $\hat{\lambda}_{\mathrm{geo}}^{(HDF)}$ evaluates the maximum metric mismatch between bounded subsets, it is exceptionally responsive to the sudden dissipation of phase space regions, effectively neutralizing the temporal lag that typically undermines standard forecast error approaches.

%----------------------------------------------------
%%%%%%%%%%%%%%%%%%%%%%%% MCC %%%%%%%%%%%%%%%%%%%%%%%%
%----------------------------------------------------
The MCC is computed as:
\begin{equation}
    \mathrm{MCC}^{(m)}_k = \frac{TP \cdot TN - FP \cdot FN}{\sqrt{(TP+FP)(TP+FN)(TN+FP)(TN+FN)}},
    \label{eq:mcc}
\end{equation}
where $k$ denotes the specific dataset and $m$ denotes the specific method, while the binary labels are assigned by the stability threshold ($\lambda_{\mathrm{thresh}}$), where a window is labeled \emph{chaotic} if $\hat{\lambda}(t) > \lambda_{\mathrm{thresh}}$ and \emph{non-chaotic} otherwise. 

Complementing the continuous tracking analysis, the binary classification results are evaluated via the $\mathrm{MCC}^{(m)}_k$ by Eq.~(\ref{eq:mcc}) are depicted in Fig.~\ref{metric_mcc_vs_qr}, demonstrate that all evaluated estimators are highly proficient at identifying the macroscopic collapse of the chaotic attractor. Across all three transient datasets, every estimator achieves an $\mathrm{MCC} \ge 0.91$. In the gradual damping scenario ($\mathcal{D}_1$), the predictive ML-FTLE baseline ($\hat{\lambda}_{\mathrm{ML}}$) establishes a marginal performance edge, with geometry-guided SSIM ($\hat{\lambda}_{\mathrm{geo}}^{(SSIM)}$) serving as the most accurate topological counterpart ($\mathrm{MCC}^{(SSIM)}_{\mathcal{D}_1} = 0.95$). Crucially, during the abrupt chaotic transitions characterizing $\mathcal{D}_2$ and $\mathcal{D}_3$, the predictive baseline and all four geometric-guided FTLE exhibit quantitative consensus, achieving identical MCC scores of $0.96$ and $0.91$ respectively. Ultimately, these high scores confirm that both the temporal prediction and spatial morphological frameworks cross the stability threshold ($\lambda_{\mathrm{thresh}}$) synchronously with the underlying dynamical transition. 

%%%%%%%%%%%%%%%%%%%%%%%%%%%%%%%%%%%%%%%%%%%%%%%%%%%%%%%%%%
%%%%%%%%%%%%%%%%%%Noise Robustness%%%%%%%%%%%%%%%%%%%%
%%%%%%%%%%%%%%%%%%%%%%%%%%%%%%%%%%%%%%%%%%%%%%%%%%%%%%%%%%
\subsection{Noise Robustness of FTLE Estimators}
To evaluate the estimators robustness against stochastic perturbations, the synthetic observables $\mathcal{D}_1$, $\mathcal{D}_2$, and $\mathcal{D}_3$ were subjected to additive white Gaussian noise (AWGN). The noise variance was scaled to simulate varying degrees of experimental measurement uncertainty, spanning signal-to-noise ratios (SNRs) from the purely deterministic limit ($\infty$ dB) to a heavily stochastic regime (0 dB). To suppress finite-sample statistical fluctuations and isolate the structural response of the estimators, all fidelity metrics were ensemble-averaged over 30 independent noise realizations per SNR level.

\begin{figure*}
    \centering
    \includegraphics[width=\textwidth]{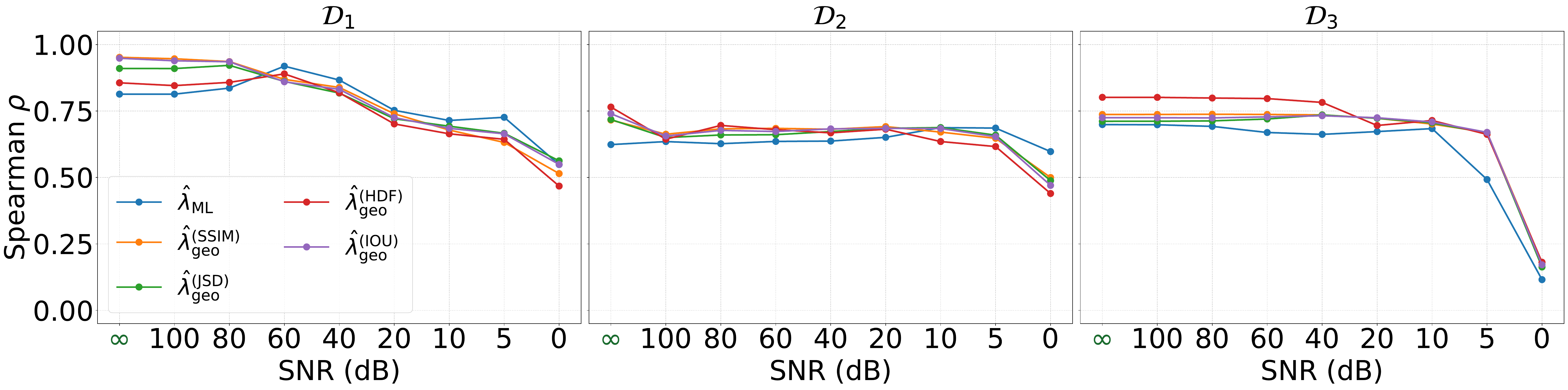}
    \caption{Spearman rank correlation ($\rho$) under additive white Gaussian noise (AWGN) for datasets $\mathcal{D}_1$, $\mathcal{D}_2$, and $\mathcal{D}_3$. The geometry-guided proxies generally preserve monotonic agreement with the Lyapunov spectrum better than the predictive ML-FTLE baseline in the moderate noise regime.}
    \label{Noise_rho}
\end{figure*}

The continuous tracking accuracy quantified by the Spearman rank correlation ($\rho$) exhibits a predictable dataset-dependent decay as stochastic entropy increases, reflecting the progressive degradation of local phase space predictability. As illustrated in Fig.~\ref{Noise_rho} the tracking performance in the gradually damped system $\mathcal{D}_1$ degrades from $\rho \approx 0.813$ at the deterministic limit to $0.548$ at 0 dB for the ML- FTLE ($\hat{\lambda}_{\mathrm{ML}}$). Notably the Geometry-guided FTLE $\hat{\lambda}_{\mathrm{geo}}^{(\mathrm{SSIM})}$ maintains a superior correlation manifold of $\rho \approx 0.952$ under low noise conditions, such as 60 dB before separating at lower signal-to-noise ratios. The abrupt topological collapse in the dataset $\mathcal{D}_2$ proves structurally resilient, where the ML-FTLE decays from $\rho = 0.624$ to $0.597$ while the geometry-guided proxy $\hat{\lambda}_{\mathrm{geo}}^{(\mathrm{IOU})}$ exhibits a high Spearman correlation of $\rho = 0.740$ and consistently outperforms the baseline across low to moderate signal-to-noise ratios such as 10 dB. In contrast, the dataset $\mathcal{D}_3$ exhibits extreme sensitivity in the low SNR limit, where the ML-FTLE collapses from $\rho=0.699$ to $0.116$ at 0 dB, indicating that severe noise completely suppresses the deterministic divergence signature associated with its transition to a periodic orbit. Meanwhile the Geometry-guided FTLE $\hat{\lambda}_{\mathrm{geo}}^{(\mathrm{HDF})}$ preserves a highly robust tracking performance, shifting only from $\rho=0.801$ to $0.714$ at 10 dB demonstrating significantly better consistency than the ML-FTLE baseline.

\begin{figure*}
    \centering    
    \includegraphics[width=\textwidth]{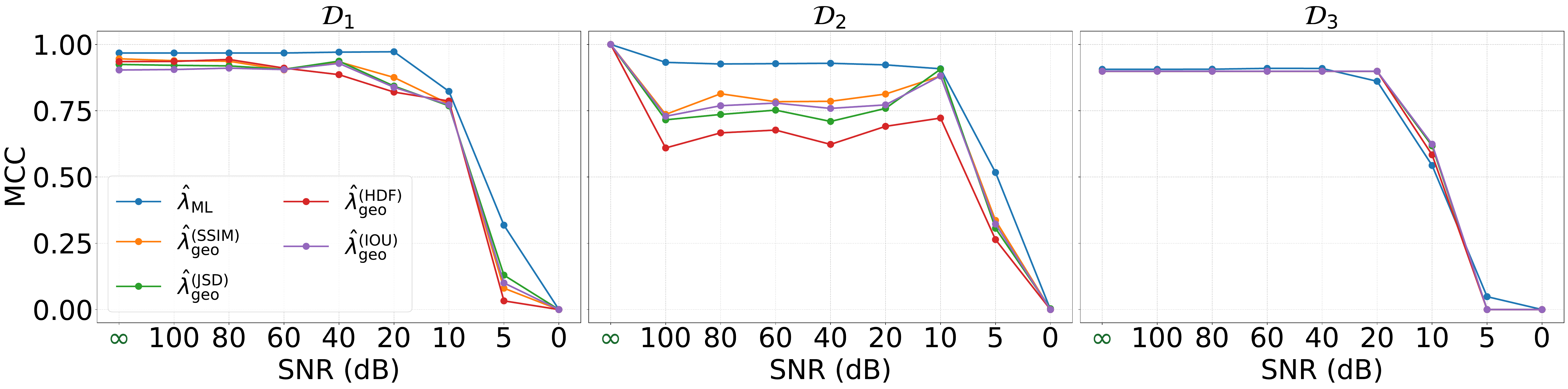}
    \caption{Matthews Correlation Coefficient (MCC) under additive white Gaussian noise for datasets $\mathcal{D}_1$, $\mathcal{D}_2$, and $\mathcal{D}_3$. Macroscopic classification remains robust down to moderate signal-to-noise ratios but collapses sharply as stochastic variance dominates the deterministic attractor topology. Results are ensemble averaged over 30 independent trials per noise level.}
    \label{Noise_mcc}
\end{figure*}

The binary classification of the chaotic to periodic and stable regime shift evaluated under this additive noise via the Matthews Correlation Coefficient (MCC) further delineates the structural breakdown illustrated in Fig.~\ref{Noise_mcc}. In the absence of additive noise, classification is exceptionally accurate across all dynamical scenarios, yielding MCC values of 0.968 for dataset $\mathcal{D}_1$, 1.000 for dataset $\mathcal{D}_2$, and 0.906 for dataset $\mathcal{D}_3$. The predictive $\hat{\lambda}_{\mathrm{ML}}$  is strongly supported by the four geometric-guided estimators, which demonstrate highly robust threshold crossing accuracy. Specifically, the ($\hat{\lambda}_{\mathrm{geo}}^{(SSIM)}$) and ($\hat{\lambda}_{\mathrm{geo}}^{(IOU)}$) geometric proxies perform exceptionally well, maintaining nearly identical classification performance to the baseline in the dataset $\mathcal{D}_1$ and $\mathcal{D}_3$ down to moderate noise levels of 20 dB where scores remain at or above 0.813 across all datasets. In dataset $\mathcal{D}_2$, while the predictive estimator maintains a higher absolute score, the geometric proxies exhibit a highly resilient and stable classification plateau across the intermediate noise spectrum. However, as stochastic perturbations intensify beyond this critical threshold, the macroscopic boundaries of the chaotic attractor become increasingly indistinguishable. At 5 dB, the MCC drops precipitously across all evaluated methods down to 0.318 for dataset $\mathcal{D}_1$, 0.517 for dataset $\mathcal{D}_2$, and 0.049 for $\mathcal{D}_3$, ultimately collapsing strictly to 0.00 at 0 dB. 

This visualizes a critical deduction, while macroscopic spatial discretization acts as an effective topological regularizer that preserves deterministic signatures under moderate corruption, extreme stochastic variance eventually completely obscures the underlying chaotic invariant measure, while direct $k$-NN forecasting relies on microscopic scalar neighborhood queries, which are rapidly dispersed by stochastic diffusion. The spatial binning inherent to the Poincar\'{e} representation acts as a macroscopic geometric spatial regularizer. This topological coarse-graining allows the Geometry-guided FTLE to retain deterministic structural information deeper into the noise floor. However, below the critical SNR threshold, the additive noise amplitude exceeds the scale of the deterministic attractor, rendering both local trajectory divergence and global phase-space morphology theoretically unrecoverable as expected due to the high stochastic rather than deterministic nature.

%%%%%%%%%%%%%%%%%%%%%%%%%%%%%%%%%%%%%%%%%%%%%%%%%%%%%
%%%%%%%%%%%%%%%--- Conclusion ---%%%%%%%%%%%%%%%%%%%%
%%%%%%%%%%%%%%%%%%%%%%%%%%%%%%%%%%%%%%%%%%%%%%%%%%%%%
\section{Discussion and Conclusion \label{conclusion}}
In this study we introduced a unified Poincar\'{e}-based geometry-guided ML-FTLE framework to address the fundamental challenge of tracking transient chaos and abrupt regime transitions solely from a univariate time series. Operating without prior knowledge of the governing differential equations, our approach successfully fuses predictive trajectory divergence and macroscopic attractor geometry. The ML-FTLE estimator extracts a local instability scale by evaluating the OLS slope of the log-transformed GMAE from out-of-sample $k$-NN forecasts. Concurrently, the framework captures the structural evolution of the reconstructed phase space by mapping delay coordinate occupancy grids onto a compact dictionary of basis attractors. By employing PLSR as a supervised latent variable bridge, we projected this high-dimensional morphological data into a calibrated Lyapunov scale.

Validation against analytical QR-FTLE baselines across diverse transient chaos scenarios confirms the high precision of this dual framework. The purely data-driven ML-FTLE reliably captures macroscopic regime shifts, distinguishing transient chaos from stable equilibria and periodic orbits. Furthermore, quantitative analysis reveals that the geometry-guided projections systematically improve continuous temporal tracking over the predictive model. Specifically, the SSIM ($\hat{\lambda}_{\mathrm{geo}}^{(\mathrm{SSIM})}$) emerges as the optimal method for smooth gradual transitions, while the HDF ($\hat{\lambda}_{\mathrm{geo}}^{(\mathrm{HDF})}$) proves highly resilient during severe topological collapses. All evaluated estimators demonstrated exceptional binary classification accuracy, achieving MCC scores $\ge 0.91$ synchronously with the underlying dynamical transitions.

The framework also demonstrates notable robustness against stochastic perturbations. Under additive white Gaussian noise, the spatial binning inherent to the Poincar\'{e} representation functions as a macroscopic geometric spatial regularizer, enabling the topological proxies to retain deterministic structural patterns deeper into the noise floor than microscopic scalar neighborhood queries. While classification performance systematically collapses at extreme noise limits such as 0 dB, where stochastic variance fully suppresses the chaotic invariant measure, the framework remains highly effective down to moderate signal-to-noise ratios of 20 dB.

Ultimately this work establishes that reconstructed phase space topology serves as a direct physical representation of dynamical structure rather than a mere visualization tool. The fusion of forecast error divergence and basis attractor morphology yields an interpretable and robust indicator of regime collapse, providing a purely empirical diagnostic tool for analyzing complex systems where only partial and noisy scalar measurements are accessible. This adaptability allows the methodology to extend into empirical domains such as climate physics and financial market analysis, where detecting abrupt structural shifts is essential.

While demonstrating exceptional reliability across synthetic benchmarks, the framework contains distinct limitations. Spatial binning and nearest neighbor searches introduce computational scaling challenges for systems with exceptionally high intrinsic dimensions, and the accuracy of the geometric proxies fundamentally depends on the quality of the initial time delay embedding. Future work will focus on processing multivariate time series natively and integrating advanced manifold learning architectures to replace uniform occupancy grids with continuous topological representations. Furthermore, implementing adaptive windowing algorithms will refine the temporal resolution for rapid dynamical collapses, ensuring the framework remains robust for real-time experimental applications.

\begin{acknowledgments}
The work of S.V.M. was catalyzed and financially supported by the Tamil Nadu State Council for Science and Technology (TNSCST), Department of Higher Education, Government of Tamil Nadu, India. %This support covered the integrated computational implementation, numerical experiments, and manuscript preparation.
I. M. was supported by the TNSCST - Science and Technology Project, Department of Higher Education, Government of Tamil Nadu, India.
A.V. was supported by the Russian Science Foundation (\href{https://rscf.ru/en/project/2211-00055/}{Grant No. 22-11-00055-P}, accessed on 10 June 2025). %The RSF support specifically covered A.V.'s contribution to the KNN forecast-divergence/GMAE-based ML-FTLE methodology, the Poincar\'{e} basis-attractor decomposition, and the interpretation of geometry-guided Lyapunov-scale instability indicators.
\end{acknowledgments}

\section*{Data Availability Statement}
The data and code that support the findings of this study are openly available on GitHub~\cite{manivelan_github_2026} and persistently archived on Zenodo~\cite{manivelan_sv_2026_20415543}.

\section*{References}
\bibliography{bibliography,refs}
\end{document}